\documentclass[12pt]{iopart}
\usepackage[utf8]{inputenc}
\usepackage{color,graphicx}
\usepackage{cite}
\usepackage{bm}
\usepackage{hyperref}
\usepackage{dsfont}

\newcommand{\bra}[1]{\ensuremath{\left\langle#1\right|}}
\newcommand{\ket}[1]{\ensuremath{\left|#1\right\rangle}}

\usepackage{ulem}  
\begin{document}
\title[Localization control under balanced pumping]{Localization control of few-photon states in parity-symmetric photonic molecules under balanced pumping}

\author{C~D~B~Bentley$^1$, A~Celestino$^1$, A~M~Yacomotti$^2$, R~El-Ganainy$^{3}$ and A~Eisfeld$^1$}
\address{$^1$ Max Planck Institute for the Physics of Complex Systems, N\"othnitzer Strasse 38, 01187 Dresden, Germany}
\address{$^2$ Laboratoire de Photonique et de Nanostructures (CNRS UPR 20), Route de Nozay, Marcoussis 91460, France}
\address{$^3$ Department of Physics and Henes Center for Quantum Phenomena, Michigan Technological University, Houghton, Michigan, 49931, USA}
\ead{cbentley@pks.mpg.de}

\begin{abstract}
	
We theoretically investigate the problem of localization control of few-photon states in driven-dissipative parity-symmetric photonic molecules. We show that a quantum feedback loop can utilize the information of the spontaneously-emitted photons from each cavity to induce asymmetric photon population in the system, while maintaining a balanced pump that respects parity symmetry. To better understand the system's behaviour, we characterize the degree of asymmetry as a function of the coupling between the two optical cavities. Contrary to intuitive expectations, we find that in some regimes the coupling can enhance the population asymmetry. We also show that these results are robust against experimental imperfections and limitations such as detection efficiency.

\end{abstract}

\maketitle

\section{Introduction} 

Spontaneous symmetry breaking (SSB) in quantum field theories is a mechanism whereby the ground state breaks one of the symmetries of its Hamiltonian. 
In many other contexts, SSB is defined broadly as the existence of a solution  (not necessarily of lowest energy) that does not respect a symmetry associated with the underlying equations of motion; it often arises due to a nonlinear-induced transition between symmetric and symmetry-broken phases. 
In systems with many degrees of freedom, these symmetry breaking instabilities can lead to several intriguing phenomena such as spontaneous pattern formation (periodic modulations, spirals, vortices, etc) and soliton wave propagation. 
These effects can be observed in a variety of physical contexts such as fluids dynamics \cite{SSB_Fluid_Dynamics}, chemical reactions \cite{chemistry_nonlinear}, Bose-Einstein condensates (BECs) \cite{SSB_BEC} and exciton-polariton systems \cite{Quantum_fluid_light}.

One of the most important platforms for investigating different symmetry paradigms is classical nonlinear optics. There SSB in the form of modulation instabilities, formation of different localized solutions (bright/dark/vortex solitons, breathers, etc.) has been already observed and investigated thoroughly in both continuous \cite{kivshar_optical_2003} and discrete setups \cite{christodoulides_discretizing_2003,kevrekidis_spontaneous_2005}. 
A minimal model for classical SSB in optics consists of two interacting modes under nonlinear conditions. Among several possibilities, this can be implemented for example by using Kerr nonlinearity in either: (1) two identical, linearly coupled waveguides (or cavities under external driving) \cite{christodoulides_discretizing_2003}; or (2) single micro-ring resonators supporting two nonlinearly coupled (via cross-phase modulation) clockwise and counter-clockwise optical modes. 
In the first system, at low power levels, the symmetric solution (having equal intensity in both waveguides or cavities) is stable. As the power input level (or the driving pump in the case of cavities) is increased beyond a certain threshold, this rather symmetric state becomes unstable and the nonlinear self-trapping effects will amplify any small symmetry-breaking perturbation. Thus in this regime, the nonlinear stable eigenmodes exhibit asymmetric intensity distribution (more localization in one waveguide versus the other). In the micro-ring arrangements, the situation is rather similar with the instability breaking the chirality of the states (see \cite{SSB_Microring_LiGe,SSB_Microring_2017} for recent experimental demonstration).

While SSB and multistabilities in  nonlinear optical setups are well-studied, a more subtle situation arises when considering the quantum regime of such systems. Particularly, the linearity of quantum mechanics (interaction terms in the Hamiltonian still lead to linear terms in the Fock space description), together with pronounced quantum noise can blur the SSB effects. To further complicate matters, until recently controlled experiments that can probe this regime~\cite{Baum10N,SSB_Exp_Yacomotti,Tren16NP} were lacking.

Recently, however, an experimental work aiming at bridging this gap has demonstrated SSB in two coupled photonic crystal laser nanocavities with only 150 intracavity photons \cite{SSB_Exp_Yacomotti}.  Subsequently, two theoretical studies modeled a coherently driven version of this experimental setup using driven-dissipative two-site Bose-Hubbard Hamiltonians \cite{SSB_Cuti,SSB_Hafezi}. These works showed that weak signatures of SSB can be traced in the quantum regime with few photons ($1 \sim 30$).
These experimental and theoretical results indicate that one will soon reach a regime where quantum effects will play a role in SSB experiments.
In these studies the driving was chosen to be time-independent.

\begin{figure}[tbp]
\centering
\includegraphics[width=8cm]{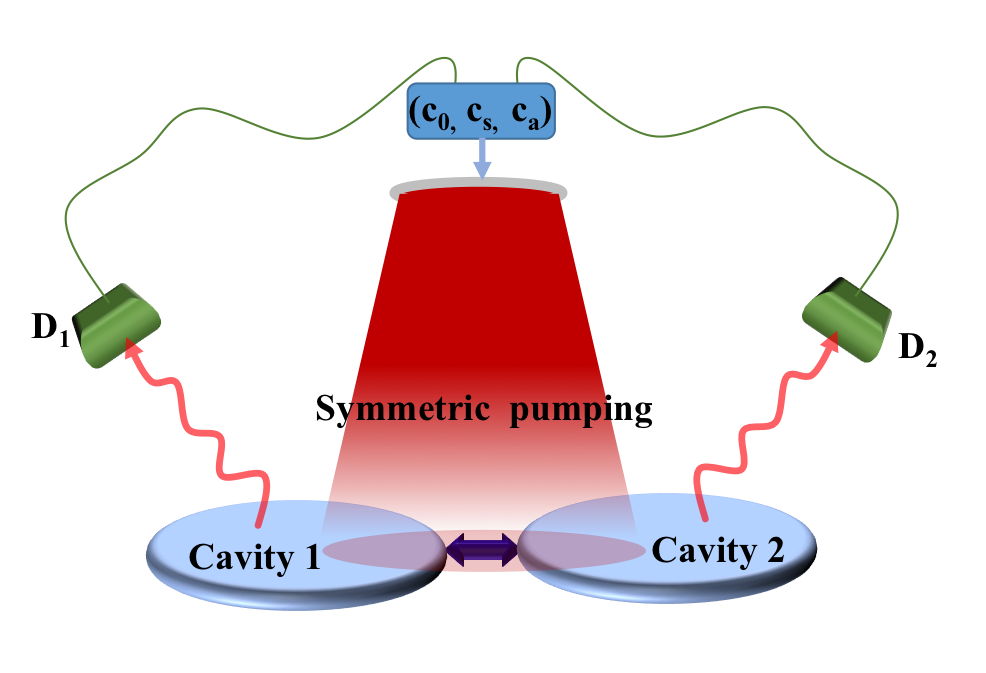}
\caption{The schematic for the considered setup: two coupled cavities with spontaneous emission measured using photodetectors.  The laser jointly drives both cavities with a state-dependent Rabi frequency. The photodetectors D$_1$ and D$_2$ are monitoring the losses of cavity 1 and 2, respectively. The input of these detectors is fed into a device, which uses this input to determine the strength of the pumping. }
\label{fig:schematic}
\end{figure}

In this paper, we consider a similar system made of {\it identical} coupled optical cavities under time-dependent {\it parity-symmetric} coherent driving and investigate the following question: can one control the degree of statistical asymmetry (as defined by the time-averaged photon population imbalance between the two cavities) without breaking the mirror-symmetry of the driving (see figure~\ref{fig:schematic})? Counterintuitively, we show that this can be achieved by using quantum feedback control~\cite{WiseQMC,Hand05JOB} with temporal pump modulation. It is important to note that while we allow the strength of the pump to vary with time, its profile remains mirror symmetric at any given instant.

The manuscript is organized as follows: in section~\ref{sec:System} we present the theoretical model of our setup.  We consider the driven, dissipative Bose-Hubbard dimer (BHD) as a simple model system, present our feedback scheme and provide a formal definition of our asymmetry measure.
The feedback scheme is applied to our system in section~\ref{sec:results}, where we show that feedback can produce asymmetry, and we discuss the robustness of our setup.
We present our conclusions and outlook in section~\ref{sec:conc}.

\section{The system and its equations of motion} 
\label{sec:System}

\subsection{Physical model}

The driven, dissipative Bose-Hubbard dimer model includes coherent and incoherent evolution for the two cavities.

The coherent evolution is given by the Hamiltonian $H = H_{\rm BH} + H_{\rm drive}$ with the Bose-Hubbard (BH) Hamiltonian \cite{Gers63PR} $H_{\rm BH}$ and a driving part $H_{\rm drive}$:
\begin{eqnarray}
H_{\rm BH} = \sum_{j=1}^2 ( \Delta  a_j^\dagger a_j + \frac{U}{2} a_j^\dagger a_j^\dagger a_j a_j ) - J (a_1^\dagger a_2 + a_2^\dagger a_1),  \label{eq:H_BH} \\ 
H_{\rm drive}(t) = \Omega(t)\, \sum_{j=1}^2 (a_j + a_j^\dagger). \label{eq:Hdr}
\end{eqnarray}
Here we have presented the Hamiltonians in the rotating frame with respect to the laser driving frequency.
With $a_j$ we denote the annihilation operator for cavity $j$.
The first term of equation~(\ref{eq:H_BH}) describes the detuning $\Delta = \omega_c - \omega_L$ of the (equal) cavity frequencies $\omega_c$ from the laser frequency $\omega_L$, the second term is an occupation-dependent nonlinear interaction with strength $U/2$ in each cavity, and the third term describes exchange of excitation  with strength $J$ between the cavities.
The coherent driving {\it symmetrically} addresses both cavities with real Rabi frequency $\Omega(t)$, which can be time dependent.

The incoherent evolution in the model represents out-coupling of photons from the cavity mode to other \emph{environmental} electromagnetic modes.
We describe the evolution of the state $\rho$ of the BHD using the following Lindblad master equation:
\begin{eqnarray}
\frac{\partial \rho}{\partial t} = -i [H(t),\rho] + \gamma \sum_j \mathcal{D}[a_j]\rho,
\label{eq:rho_dot}
\end{eqnarray}
where  $\mathcal{D}[a_j]\rho = a_j \rho a_j^\dagger - \frac{1}{2} \left( a_j^\dagger a_j \rho + \rho a_j^\dagger a_j \right)$ terms describe the effect of the photon outcoupling on the system. 
Throughout the work we use $\gamma$ as the unit of energy.

Since our feedback scheme, depicted in figure~\ref{fig:schematic} and discussed in detail below, leads to stochastic contributions in the evolution of $\rho$ it is convenient to use a stochastic-trajectory-based description from the outset. This is, in particular, numerically advantageous as the dimension of the state is the square root of the dimension of the density matrix\footnote{For the calculation used in the present work we typically have a basis with dimension $\sim 400$.}.
Individual trajectories are propagated with the following stochastic Schr\"{o}dinger equation (SSE)~\cite{WiseQMC}: 
\begin{eqnarray}
\mathrm{d} \ket{\psi (t)} 
= &\sum_j   \mathrm{d}\xi_j(t) \left( \frac{a_j}{\sqrt{n_j(t)}} - 1 \right) \ket{\psi (t)} \nonumber \\
&+ \mathrm{d}t  \Bigg(\frac{\gamma}{2} \sum_j  \Big(n_j(t)-  a_j^\dagger a_j \Big) -i H(t) \Bigg) \ket{\psi (t)}, \label{eq:stschr}
\end{eqnarray}
where the time dependent populations of the two sites $j=1,2$ are given by 
\begin{equation}
n_j(t)=\bra{\psi(t)}a^{\dagger}_j a_j \ket{\psi(t)}.
\end{equation} 
and the stochastic increments $d\xi_j$ have the following properties:
\begin{eqnarray}
\mathrm{d}\xi_k (t) \mathrm{d}\xi_j (t) = \mathrm{d}\xi_k (t)\, \delta_{kj} \label{eq:dxidxi}\\
\mathcal{E}[\mathrm{d}\xi_j(t)] = \gamma n_j(t) \mathrm{d}t. \label{eq:E[dxi]}
\end{eqnarray}
Here $\mathcal{E}[\cdots]$ denotes the average over trajectories.
The solution of equation~(\ref{eq:rho_dot}) is approached by taking the average of many trajectories.
The stochastic equation, equation~(\ref{eq:stschr}), was solved using the XMDS package~\cite{Denn13CPC,xmdsweb}.
Note that equation~(\ref{eq:dxidxi}) implies that the stochastic increment is either 0 or 1.

Besides being numerically more efficient, the SSE~(\ref{eq:stschr}) is also a model for certain types of measurements which will be relevant for the feedback scheme presented below. 
This equation~(\ref{eq:stschr}), with the given stochastic statistics in equations~(\ref{eq:dxidxi},~\ref{eq:E[dxi]}), describes the system evolution with photodetection of the photons emitted from each cavity, assuming perfect detection efficiency and no delay in time between photon emission and detection \cite{WiseQMC}.
A single quantum trajectory corresponds to a single run of an experiment.
The detection events correspond to the stochastic noise increment $\mathrm{d}\xi_j(t) \in \{0,1\}$ at any time $t$, for the detector at site $j$.
The first term in equation~(\ref{eq:stschr}) thus represents a photodetection event and corresponding `jump' in the state when $\mathrm{d}\xi_j=1$, and no detection when $\mathrm{d}\xi_j=0$.

We will make use of this photodetection model for our feedback scheme.
The photodetection record (history of `clicks' from the detector) provides information about the population in each cavity~\cite{Bout07SIAMJCO}.
In fact, given the photodetection record and knowledge of the Hamiltonian parameters, we have perfect knowledge of the quantum state of the system at any given time.
We can then employ properties of the state in our feedback scheme to control the state evolution.

\subsection{The feedback scheme}

In our feedback scheme the value of the driving at time $t$ is determined by the state of the system, i.e, $\Omega(t) \rightarrow \Omega \cdot f(\psi_t)$, where $\Omega$ determines the overall strength of the driving and $f(\psi_t)$ contains the time-dependence via $\psi_t \equiv \ket{\psi (t)}$.
Accordingly the driving Hamiltonian reads
\begin{equation}
H_{\rm drive} (t) \rightarrow H_{\rm drive} (\psi_t) = \Omega \cdot f(\psi_t) \sum_{j=1}^2  (a_j + a_j^\dagger).
\label{eq:Hdrt}
\end{equation}
Through this driving the Hamiltonian $H(t)$ in equation~(\ref{eq:stschr})   becomes explicitly state-dependent. 
Note that at any given time, the driving addresses both sites with the same driving strength.
In the following we set $\Omega=\gamma$. Thus all relevant information of the (time-dependent) driving strength is encoded in the feedback function $f$.

We are now interested in what effect this state-dependent driving will have on the system, and in particular, how asymmetry can be engineered.
To this end we consider a simple functional form for the driving feedback:
\begin{eqnarray}
f(\psi_t;\vec{c}) 
 = c_0 -c_{\rm s}\cdot n_{\rm tot}(t;\vec{c})- c_{\rm a}\cdot n_{\rm diff}(t; \vec{c})  
 \label{eq:def_feetback},
\end{eqnarray}
Here the time-dependent total population and the time-dependent population difference are defined as 
\begin{eqnarray}
n_{\rm tot}(t)&=&n_1(t) +n_2(t),\\
n_{\rm diff}(t)&=& n_2(t) -n_1(t).
\end{eqnarray}
The argument $\vec{c}\equiv \{c_0, c_s, c_a \} $ indicates the parametric dependence of the populations (and $f$) on the choice of the parameters $c_0$, $c_{\rm s}$, $c_{\rm a}$ of the feedback function.
These coefficients can tune the state dependence from the symmetric case $c_a = 0$ to more general asymmetric state-dependence.
For $c_a=c_s=0$, constant driving with no feedback is recovered.

\subsection{Asymmetry measure} \label{sec:asymmm}

We now introduce an asymmetry parameter to characterize the imbalance in the photon population:
	\begin{equation}
	A (\vec{c}) =   \frac{N_\mathrm{diff}(\vec{c})}{N_\mathrm{tot}(\vec{c})}
	\label{eq:Assymetry_optimizer}, 
	\end{equation}
	with
	\begin{eqnarray}
	N_\mathrm{diff}(\vec{c}) =   \frac{1}{M}\sum_{\ell=1}^M\frac{1}{T}\left[\int_{t_i}^{t_f} n^{(\ell)}_{\rm diff}(t;\vec{c}) dt\right]  \label{eq:N_diff}, \\
	N_\mathrm{tot}(\vec{c}) =   \frac{1}{M}\sum_{\ell=1}^M\frac{1}{T}\left[\int_{t_i}^{t_f} n^{(\ell)}_{\rm tot}(t;\vec{c}) dt\right]  \label{eq:N_tot} 
	\end{eqnarray}
	Here the summation over $\ell$ indicates the averaging over the time-average  of $M$  individual trajectories. The time-average is performed by integrating from $t_i$ to $t_f$.

We are primarily interested in the relative population difference. Thus, in equation~(\ref{eq:Assymetry_optimizer}), we have normalized by the average total population. Accordingly, the values of $A (\vec{c})$ range from $-1$ to $1$. If cavity 1 is higher populated than cavity 2, then $A(\vec{c})$ is negative. Note that switching the sign of $c_a$ changes the sign of $A(\vec{c})$. Thus for $c_a=0$ we expect that $A=0$, and this is indeed what we find in the numerical simulations.

Henceforth when we refer to asymmetry we refer to the definition in equation~(\ref{eq:Assymetry_optimizer}).

\section{Quantum control of few-photon localization} \label{sec:results}

\subsection{Localization results}

Now that we have defined the model and its governing equation, we return to our initial question: can one control the degree of asymmetry in a parity-symmetric setup without breaking the mirror-symmetry of the driving?

To this end, we now consider a concrete example with  $J=0.05$$\gamma$, $U=0.3$$\gamma$ and $\Delta=-1.65$ $\gamma$ (we discuss this choice of parameters in the Conclusions). 

\begin{figure}[th]
\centering
\includegraphics[width=14cm]{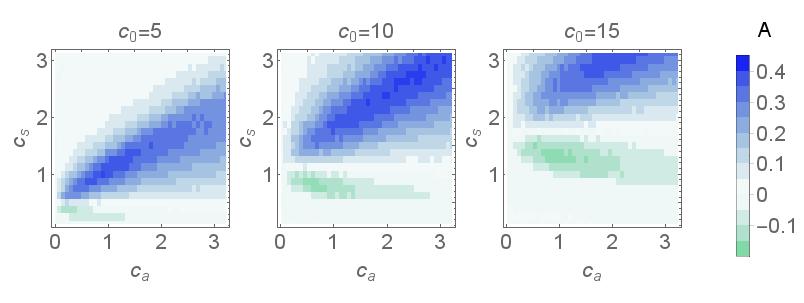} 

\caption{Asymmetry $A (\vec{c})$ with $c_{\rm a}$ and $c_{\rm s}$, for different values of $c_0$.
Here $J=0.05$$\gamma$, $U=0.3$$\gamma$ and $\Delta=-1.65$ $\gamma$.
Each point is calculated using 100 trajectories.
We estimate the error in $A (\vec{c})$ to be on the order of 0.03 using 200 additional trajectories (with duration $200/\gamma$) for each point.
Since we restrict the number of $\vec{c}$-tuples that the optimization routine tests to 30 (see appendix A), the obtained values should be considered to be approximate lower bounds for the achievable asymmetry with our feedback scheme.
}
\label{fig:asc2c3}
\end{figure}

Figure \ref{fig:asc2c3} (a)-(c) shows the asymmetry measure $A$ as function of the feedback parameters $c_{\rm a}$  and $c_{\rm s}$ for three different values of $c_0$. Each point on these figures has been generated by using 100 trajectories with a time span of $200/\gamma$.

Interestingly, even with such a crude parameter scan, one sees a clear evidence of localization control as characterized by a non-zero value of $A$.  Furthermore, for each $c_0$ there is a broad, smoothly-connected region which gives maximal asymmetry $A\approx 0.4$. This observation persists even if we repeat the simulations by varying $c_0$. We emphasize here that this occurs despite the fact that we employ balanced driving that respects parity symmetry. 
What makes this result particularly interesting is that for few photons in the system ($\sim$2-8, as seen in figure~\ref{fig:controlerrs}), quantum fluctuations play a significant role.
The driving, despite its parity symmetry, makes use of these fluctuations to generate asymmetry in the photon population.
This intriguing observation is the central result of this work.

Next, we investigate how the coupling coefficient impacts the degree of localization. From a computational perspective, finding large values for the asymmetry measure $A$ (ideally the maximum value) for a given set of parameters ($\Delta$, $U$ and $J$) requires an expensive search in the three dimensional feedback parameter space spanned by the vector $\vec{c}$. In order to overcome this difficulty, we employ the Nelder-Mead optimization scheme~\cite{NelderMead1965} (see appendix A for more details). This approach reduces the computational cost considerably while yielding large values for the asymmetry parameter $A$, which we believe to be very close to the optimal asymmetry that can be achieved (see appendix A).

\begin{figure}[tp]
\centering
\includegraphics[width=9cm]{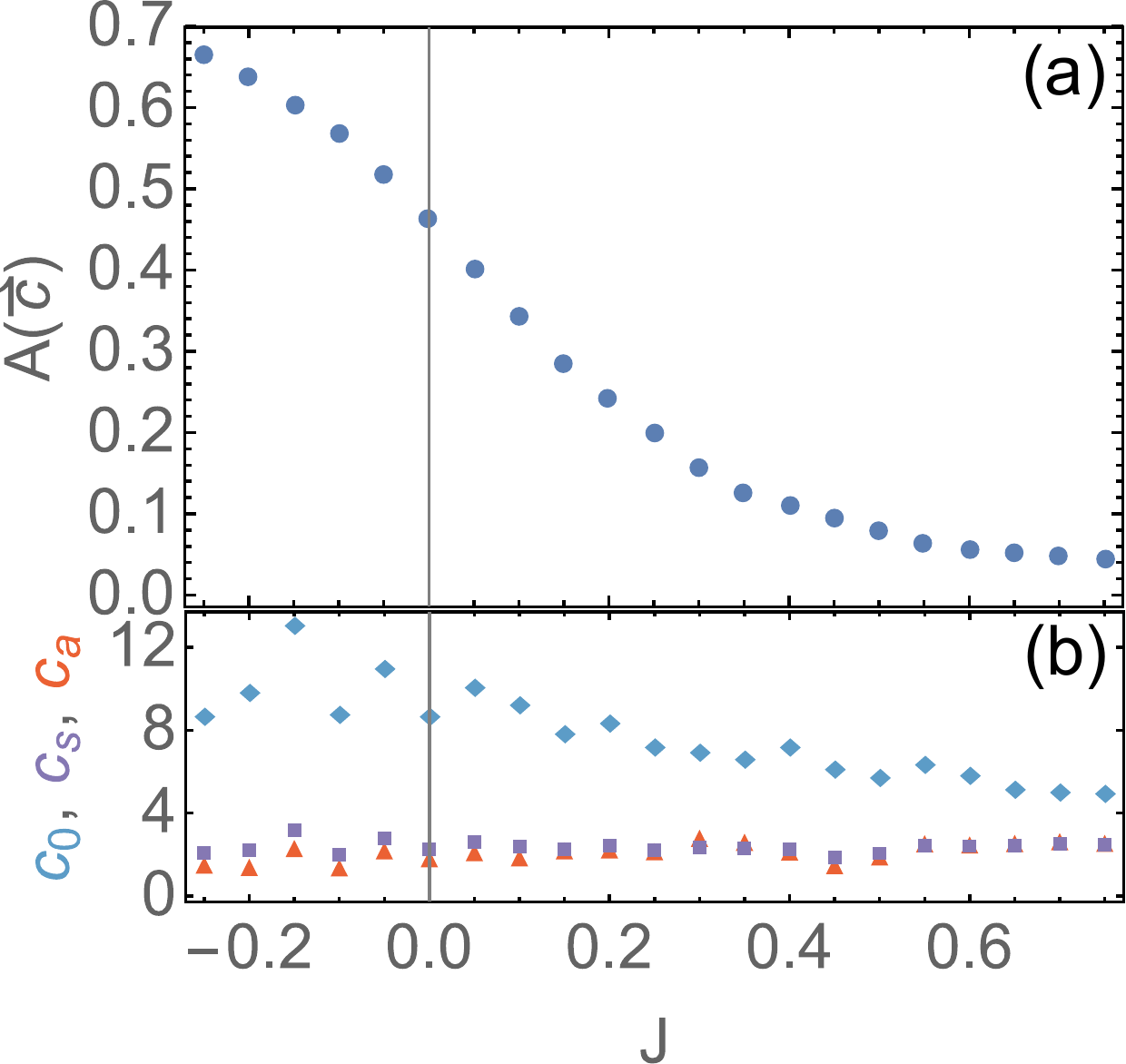}
\caption{\label{fig:3} (a) Asymmetry $A$ with feedback for $\Delta=-1.65\, \gamma$ and $U=0.3\, \gamma$ as a function of $J$. For each $J$-value the $\vec{c}$ have been optimized to yield a large asymmetry. (b) The respective parameters $\vec{c}$ from the optimization are displayed.
}
\end{figure}

Figure~\ref{fig:3} (a) plots the asymmetry measure $A$ as a function of the coupling $J$ when $\Delta=-1.65$ and $U=0.3$ (same parameters as above). Naively, one would expect that $A$ should drop as a function of $|J|$ since coupling provides a pathway for photons to move between the two cavities, making it harder for the control to favor one cavity over the other. Indeed this is the case for positive values of $J$ as can be seen in Fig. \ref{fig:3} (a). Surprisingly however, for negative values of $J$, we find that the asymmetry increases with $|J|$, approaching large values (one should keep in mind that a value of 1.0 means perfect localization on one cavity). Our simulations thus indicate that the naive picture is not always correct and that coupling can provide a pathway for increasing asymmetry.
For each value of $J$, the optimized values of the feedback parameters are shown in Fig. \ref{fig:3} (b). One sees that these parameters stay roughly constant over the entire $J$ range, the largest variation being in $c_0$. For completeness, we have also computed the values of $A$ with no feedback ($c_s=c_a=0$), where we found that  $A=0$. As expected, we also found that $A=0$ for $c_a=0$.

\subsection{Robustness of the feedback}

While the main objective of the present work is to demonstrate that via symmetric feedback one can obtain quite strong asymmetry in the populations of the cavities, it is nevertheless interesting/important to discuss the robustness of this scheme  with respect to imperfections that would always be present in an experimental realization.

\paragraph{Variation in the control parameters $\vec{c}$:}
One important aspect is how sensitively the asymmetry $A(\vec{c})$ depends on $\vec{c}$. More precisely, will small deviations from the values found from optimization still yield similar asymmetry?
To gain insight into this question, it is instructive to look at figure~\ref{fig:asc2c3}.  
There we see that the region around the maximum is quite broad and smooth.
Thus small variations around the maximum will not change the value of the asymmetry significantly.
For various $J$ we have checked that small deviations from our numerically optimized values $\vec{c}$ indeed do not change the asymmetry significantly.

\paragraph{Changes in the driving field are not infinitely fast:} 
In experiments the change in the driving cannot be arbitrarily fast. 
To take that into account we define the maximum rate of change by
\begin{equation}
r_\mathrm{max}=|\mathrm{d}f/\mathrm{d}t|.
\end{equation}
The ideal driving control varies on a timescale similar to that of the mean populations of the two sites, as can be seen from equation~(\ref{eq:def_feetback}) and in figure~\ref{fig:trajs}.
One expects that a reduction of this rate of change will lower the achievable asymmetry.

\begin{figure}[t!]
\centering
\includegraphics[width=16cm]{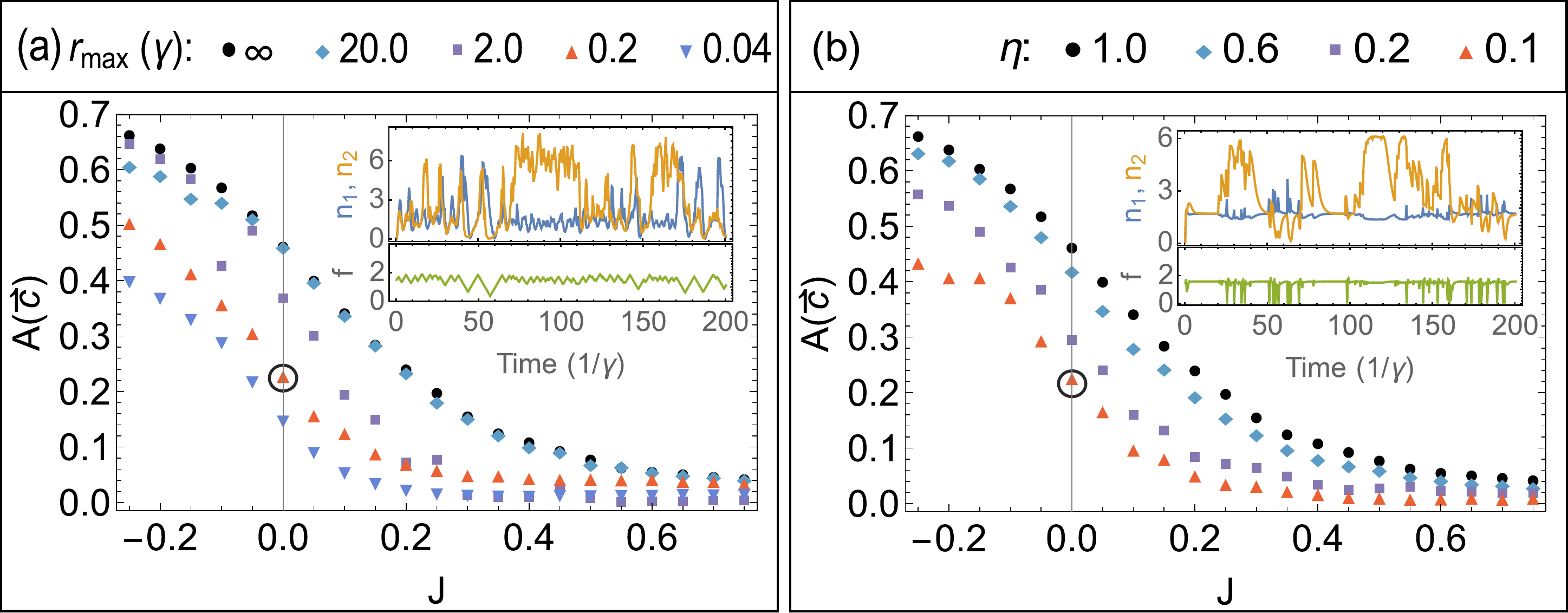}
\caption{Asymmetry with feedback for non-ideal conditions.
(a): Finite rate of change of the driving field. The asymmetry for various values of $r_{\textrm{max}}$ is shown.
Optimization over $\vec{c}$ was performed for each $r_{\mathrm{max}}$. (b): Imperfect detection efficiency. Here $\eta$ is the fraction of emitted photons detected. The optimal $\vec{c}$ for $\eta=1$ is used for each value of $\eta$.
All parameters are the same as in figure~\ref{fig:3}.
The insets show examples of single trajectories with the corresponding driving for the encircled data points.
}
\label{fig:controlerrs}
\end{figure}

The effect of limiting the rate of change $r_\mathrm{max}$ in the applied driving is demonstrated in figure~\ref{fig:controlerrs}(a).
One sees that even for a slow maximal rate compared to $\mathrm{d} n_{1,2}/\mathrm{d}t$, one can still obtain pronounced asymmetry. In the inset one nicely sees the finite rate of change in the driving function $f(t)$.

\paragraph{Detection efficiency:}
When a cavity spontaneously emits a photon, only a fraction of these spontaneous emission events is collected by the respective photodetector.
In the following, we use $\eta$ to denote the fraction of detected photons. 
The ability to detect the photons emitted from the cavities plays a crucial role in our feedback scheme. 
Thus one expects that imperfect detection efficiency ($\eta < 1$), which is always present in experiments, will strongly influence our results.
However, numerically we find that even for quite small efficiencies ($\eta\sim 0.1$) the feedback scheme works surprisingly well, as can be seen in  figure~\ref{fig:controlerrs}(b).
In this figure we present numerical calculations for the same parameters as in figure~\ref{fig:3} for different values of $\eta$.
Remarkably, we also used the same $\vec{c}$ values which are optimized for the perfect detection case. 
We suspect that even higher asymmetry could be achieved if we optimize taking the known detector efficiency into account.
It is also instructive to take a look at the inset where a single trajectory with the corresponding driving is shown for $\eta=0.1$.
Comparing to the trajectory and feedback function with $\eta=1$ (see figure~\ref{fig:trajs}) one sees that for $\eta=0.1$ the feedback is adjusted less often, as expected. 

Our numerical implementation is based on the method discussed in pages 190-191 of Ref.~\cite{WiseQMC}.

\paragraph{The Hamiltonian and the coupling to an environment are not exactly known.}
Our feedback scheme relies on the fact that we are able to infer the population of the cavities from measurements. 
The situation becomes complicated if the parameters entering the Hamiltonian are not known exactly.
One also has to keep in mind that our model Hamiltonian is always an approximation to the real experimental system.
We suspect that our feedback scheme will still work if the deviations from our ideal situation are not too large.
However, such a study (as in~\cite{Wars02PRA,Comb11JPB,Szig13PRA}) is beyond the scope of the present work. 
For each experimental setup this has to be individually investigated.

\section{Conclusions} \label{sec:conc}

In this work, we have investigated the problem of few-photon localization in a system of two mirror-symmetric coupled optical cavities modeled by a driven-dissipative Bose-Hubbard Hamiltonian with many-body interactions. We have shown that, in this regime, state-dependent quantum feedback control can be utilized to promote an asymmetric state with more photons in one cavity than the other, even when the driving itself is kept symmetric. Intuitively, this result can be better understood by reference to Maxwell's demon. In the famous problem of gas molecules confined in a container divided symmetrically by a wall having a small hole, the demon can measure the position of the molecules on both sides of the wall and use their random motion to control the opening of the hole to force the molecules to move from one side to the other. Analogously, our scheme is based on a similar principle, except that here randomness originates from quantum fluctuations rather than thermal noise.

In the manuscript, we have demonstrated localization using an exemplary set of parameters.
This parameter set is not unique.
We have found that the feedback generates localization for the following regions about these parameter values: $U/\gamma = 0.3 \pm 0.15$ and $\Delta/\gamma = -1.65 \pm 0.6$.

We now comment on the experimental feasibility of our results using optical setups. The main challenge here would be the required fast feedback timescales.
Particularly, our scheme relies on feedback in the form of continuous, time-dependent driving whose strength is modulated in time as a function of the cavity populations. Practically, this temporal modulation could be introduced by electro-optic modulators. Typical time scales of these devices are in the order of 1 ns. If we consider a system similar to that studied in
\cite{SSB_Exp_Yacomotti} and having a damping rate of $\sim 7$~ps, we find that a time scale of 0.7~ns for variation in the driving (to increase from its minimum ($f_\mathrm{min} = 0.01$) to its maximum value ($f_\mathrm{max} =4.0$)) achieves reasonable results (with $r_{\mathrm{max}}=0.04$ as discussed in figure~\ref{fig:controlerrs}). Clearly, the required time scales are not far from those provided by current on-shelf electronic components.  It is thus foreseeable that near-future electronic technologies will be adequate for use in such experiments. 

An attractive alternative would be polariton-based setups in cavity QED~\cite{Hart08LPR}, which can have a damping timescale on the order of 1~$\mu$s.  This would render a feedback delay timescale of 1~ns negligible.

\section{Acknowledgments}

R.E. acknowledges support from Henes Center for Quantum Phenomena at Michigan Technological University; and the Max Planck Institute for the Physics of Complex Systems (MPIPKS).
R.E. thanks the visitors program of the MPIPKS for making an extended stay at the MPIPKS possible during which most parts of the present work have been produced.
A.M.Y. acknowledges  support from the “Investissements d’Avenir” program (Labex NanoSaclay, Grant No. ANR-10-LABX-0035), and the ANR UNIQ DS078.
A.E. and C.D.B.B. gratefully acknowledge discussions with Francesco Piazza and Jad C Halimeh.

\clearpage
\appendix

\section{Optimization details}

As noted in section 3.1 in the main text, we want to find parameters $\vec{c}$ of the feedback function that lead to an asymmetry in the population of the two cavities.
Ideally we want the asymmetry measure $A(\vec{c})$ to be as large as possible (within reasonable values of the driving function $f(t;\vec{c})$).
Since we do not have an analytic relationship between $A(\vec{c})$ and $\vec{c}$ (for any set of parameters $J$, $\Delta$ and $U$), we numerically calculate $A(\vec{c})$ for different parameter vectors $\vec{c}$ (and fixed $J$, $\Delta$ and $U$).
Scanning the parameter space of $\vec{c}$ by numerically computing $A(\vec{c})$ (as in figure~\ref{fig:asc2c3}) is too computationally costly to perform. 
Therefore our goal is to find as large an $A(\vec{c})$ as possible, while testing as few $\vec{c}$ samples as possible.

For this task we use the Nelder-Mead optimization algorithm~\cite{NelderMead1965}, as implemented in Mathematica. 
In this algorithm a cost function $f_\mathrm{cost}(\vec{c})$ is minimized.
In our case the cost function is $-A(\vec{c})$.
For practical reasons, during the optimization process we numerically evaluate this expression slightly differently than described in the main text (see equation~(\ref{eq:Assymetry_optimizer})).
We calculate the cost function as $f_\mathrm{cost}(\vec{c})=-A_\mathrm{opt}(\vec{c})$ with
\begin{equation}
A_\mathrm{opt}(\vec{c})  = \frac{1}{N_a} \sum_{j=1}^{N_a} A_j (\vec{c}).
\end{equation}
where 
\begin{equation}
A_j(\vec{c})= \frac{\sum_{k=1}^{N_T/N_a} \int_{t_i}^{t_f} n_\mathrm{diff}^{j,k} (t,\vec{c}) dt}{\sum_{k=1}^{N_T/N_a} \int_{t_i}^{t_f} n_\mathrm{tot}^{j,k} (t,\vec{c}) dt}   
\end{equation}
Here $N_\mathrm{T}$ is the total number of trajectories and $N_\mathrm{a}$ is the number of 'sub-ensembles' in which we group these trajectories.
For each sub-ensemble $j$ the asymmetry $A_j$ is calculated in accordance with the description in the main text (see equation~(\ref{eq:Assymetry_optimizer}) and the discussion afterwards). 
That means for each trajectory $k$ from sub-ensemble $j$ we integrate $n_\mathrm{diff}^{j,k}$ and $n_\mathrm{tot}^{j,k}$ from time $t_\mathrm{i}$ to $t_\mathrm{f}$ which are the same for each trajectory.
In the simulations shown we have used $t_\mathrm{i}=0.4$ and $t_\mathrm{f}=200$ (in units of $\gamma$).
The choice $t_\mathrm{i}=0.4$ removes the short initial transients.

For $N_\mathrm{a}=1$ we have $A_\mathrm{opt}=A$ from equation~(\ref{eq:Assymetry_optimizer}).
If the ratio $N_\mathrm{T}/N_\mathrm{a}$ is large then also $A_\mathrm{opt}\approx A$.
For the values we have used ($N_\mathrm{T}=1000$ and $N_\mathrm{a}=10$) we have found that  $A_\mathrm{opt}(\vec{c})$ is almost identical to $A(\vec{c})$.

Since the different $A_j(\vec{c})$ are calculated from independent trajectories, their values will be different. 
We use the spread of these values to estimate the accuracy of  $A_\mathrm{opt}(\vec{c})$. 
To this end we consider the standard deviation
\begin{equation}
\sigma_{A_\mathrm{opt}}=\sqrt{\frac{1}{N_\mathrm{a}-1}\sum_{j=1}^{N_\mathrm{a}} (A_j-A_\mathrm{opt})^2}
\end{equation}

Clearly, by increasing the total number of trajectories $N_\mathrm{T}$ the accuracy will increase. 
For our choice of $t_\mathrm{i}$, $t_\mathrm{f}$, $N_\mathrm{T}$, and $N_\mathrm{a}$ we find $\sigma_{A_\mathrm{opt}}$ values on the order of  0.003.

All results presented are calculated using $N_\mathrm{S}=30$ iterations of the optimization algorithms; that means $A_\mathrm{opt}(\vec{c})$ is evaluated for  $N_\mathrm{S}$ different choices of $\vec{c}$.
The $\vec{c}$ which provides the largest value is then used to calculate $A(\vec{c})$.
The optimized $\vec{c}$ parameters used in figure 3 are shown in table~\ref{tab:cparams}.
The initial parameter ranges for $\vec{c}$ roughly correspond to the ranges shown in figure 2 of the manuscript.

We believe that our optimization routine produces results close to the global maximum for the asymmetry.
Firstly, the parameter scan in figure 2 for $J=0.05 \gamma$ shows a maximal asymmetry region.
Our optimization routine produces feedback parameters corresponding to this region, with a matching asymmetry value, within the given accuracy for figure 2.
The feedback parameters are given in table~\ref{tab:cparams} and plotted in figure 3.
Secondly, repeated or additional optimization runs have produced consistent results.

\begin{table*}[tb]
\caption{\label{tab:cparams} Table of optimized feedback coefficients used in figure 3.}
\centering
\begin{tabular}{cccc}
J & $c_0$ & $c_s$ & $c_a$\\ \hline
-0.25 & 8.61 & 1.89 & 1.46 \\
-0.20 & 9.76 & 2.01 & 1.35\\
-0.15 & 12.99 & 2.98 & 2.26\\
-0.10 & 8.70 & 1.80 & 1.32\\
-0.05 & 10.91 & 2.59 & 2.15\\
0. & 8.61 & 2.07 & 1.77\\
0.05 & 10.01 & 2.43 & 2.06\\
0.10 & 9.16 & 2.21 & 1.81\\
0.15 & 7.77 & 2.07 & 2.15\\
0.20 & 8.28 & 2.22 & 2.20\\
0.25 & 7.13 & 2.02 & 2.11\\
0.30 & 6.88 & 2.14 & 2.75\\
0.35 & 6.54 & 2.10 & 2.56\\
0.40 & 7.13 & 2.07 & 2.09\\
0.45 & 6.06 & 1.69 & 1.43\\
0.50 & 5.66 & 1.87 & 1.85\\
0.55 & 6.29 & 2.22 & 2.49\\
0.60 & 5.76 & 2.22 & 2.43\\
0.65 & 5.09 & 2.24 & 2.50\\
0.70 & 4.96 & 2.33 & 2.59\\
0.75 & 4.89 & 2.29 & 2.51\\
\end{tabular}
\end{table*}

\section{Quantum trajectories}

Here, we outline the calculation of the system dynamics, and present an illustrative example.

\subsection{Calculation of trajectories}
Individual trajectories are calculated by propagating the state vector $\ket{\psi(t)}$ according to the SSE~(\ref{eq:stschr}).
Both cavities are initially empty, before pumping populates them with photons.
The trajectories become different because of the stochastic events corresponding to photon emission.
Through the feedback $f(\psi_t,\vec{c})$ (from equation~(\ref{eq:def_feetback})) the evolution of the state also depends parametrically on the choice of $\vec{c}$.

For numerical reasons, we impose the bounds $f_\mathrm{min} \leq f(\psi_t,\vec{c}) \leq f_\mathrm{max}$, with $f_\mathrm{min}=0.01$ and $f_\mathrm{max}=4.0$.
If equation~(\ref{eq:def_feetback}) gives a value of $f(t)$ larger (smaller) than $f_\mathrm{max}$ ($f_\mathrm{min}$) then $f(t)$ is set to be $f_\mathrm{max}$ ($f_\mathrm{min}$).
The lower bound prevents the cavities from remaining empty.
The upper bound prevents occupation of very high photon-number states in either cavity.

For the parameters that we consider, we can truncate the Hilbert space of the state vector to less than 20 photons in each cavity. Thus the dimension of the Hilbert space is $20 \times 20 = 400$.

\subsection{Example trajectories}

\begin{figure*}[t!]
	\centering
	\includegraphics[width=16cm]{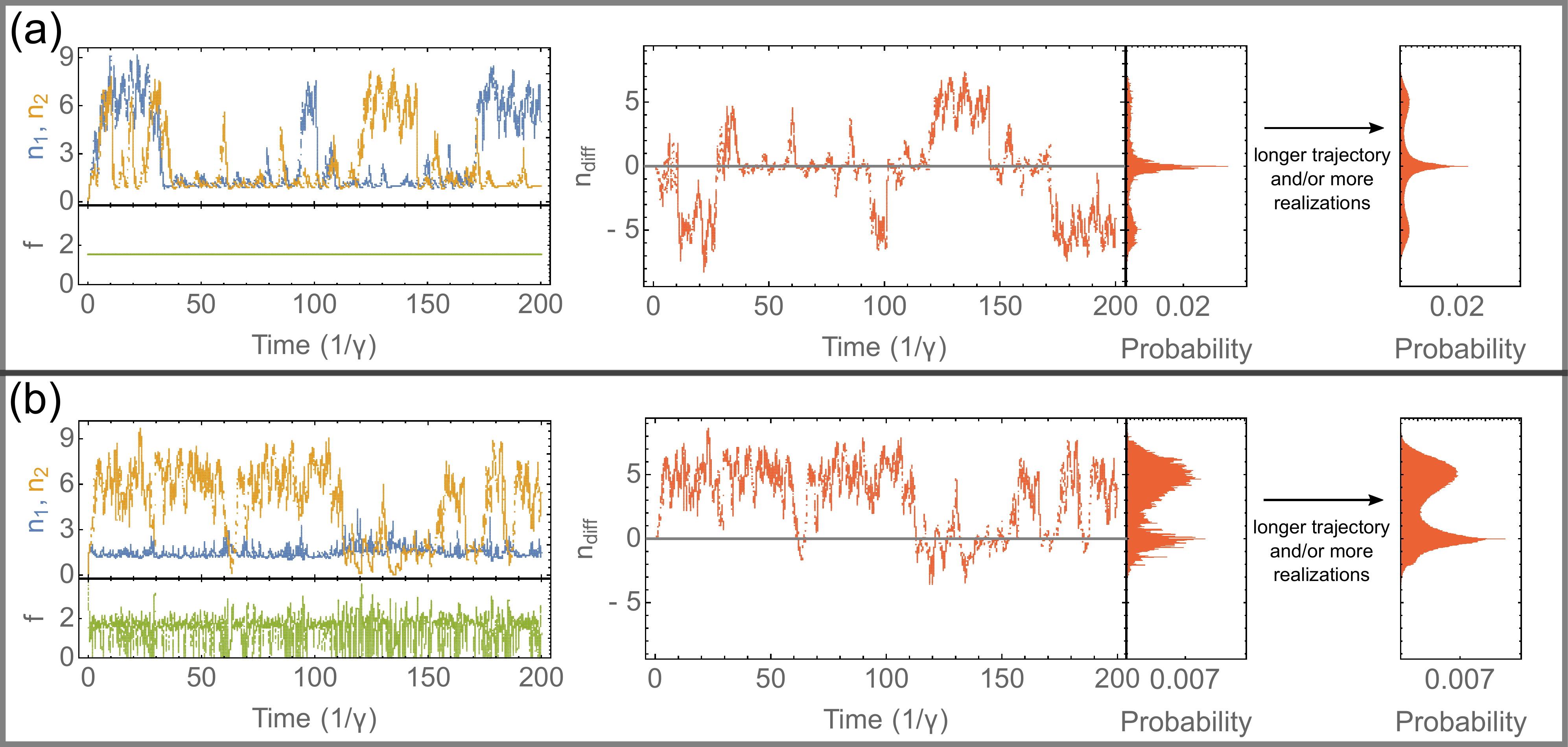}
	\caption{
		Trajectories of the mean cavity populations for a particular photodetection record are shown for {\it no} feedback (a) and {\it with}  feedback (b). 
		The 'feedback functions' $f (t)$ are also shown (green curves). In (a) it has a time constant value $f(t)=1.5$. 
		The population difference $n_{\rm diff}(t)$ is shown in the middle panel with the probability to find a certain value of $n_{\rm diff}$ in the given trajectory. 
		In the right panel, the histogram $p(n_\mathrm{diff})$ is shown, which is obtained by averaging over many trajectories (we obtain the same result if we consider a very long trajectory).		
		Here, $U=0.3 \gamma$, $\Delta = -1.65 \gamma$ and  $J=0.0 $.
		As an example of the asymmetry measure $A$ (equation~(\ref{eq:Assymetry_optimizer})) we find $A=0.00$ and $A=0.46$ for (a) and (b) respectively.
		}
		\label{fig:trajs}
		\end{figure*}
		
		To demonstrate the effect of the feedback we consider the case of uncoupled cavities, i.e., $J=0$.
		For the detuning and the nonlinearity we choose $\Delta =-1.65 \gamma$, and $U=0.3 \gamma$, respectively.
		In figure~\ref{fig:trajs}, we illustrate the difference between quantum trajectories with constant driving (a) and with feedback (b). 
		For the constant driving case (a) we have used $f(t)= 1.5 ={\rm const}$. 
		In (b) we have used $c_0=8.61$, $c_{\rm s}=2.07$ and $c_{\rm a}=1.77$.
		This choice of parameters resulted in a feedback function $f(t)$ that has the same mean $\int_0^T {\rm d}t f(t)/T=1.5$ as the constant driving case (a).

		In the left panel results of single trajectories are displayed and $n_1$ and $n_2$ are shown together with the feedback function.
		The initial state was both cavities empty, i.e.~$n_1(0)=n_2(0)=0$.
		In the middle panel the population difference $n_{\rm diff}(t)$ is shown together with the probability to find a certain value of $n_{\rm diff}$ in the respective trajectory. 
		Finally in the right panel we show the corresponding histogram $p(n_\mathrm{diff})$, obtained by averaging over many trajectories (we obtain the same result if we consider a longer trajectory).
		Here we have also removed the first 0.4 time-units during which the initial transient dynamics takes place\footnote{We observed very little influence on our results when taking this initial dynamics into account.}.
		
		For the case without feedback (a), most of the time the cavities have equal populations, resulting in a strong peak at $n_{\rm diff}=0$ in the histogram.
		Weak shoulders can be seen symmetrically around $n_{\rm diff}=0$, indicating that the system sometimes does not have the same population in the two cavities.
		Note that the symmetry means that the probability that cavity 2 has more photons than cavity 1 ($n_{\rm diff}>0$) is the same as the probability that cavity 1 has more photons than cavity 2 ($n_{\rm diff}>0$). 
		
		In the case with feedback (b), one clearly sees in the trajectory shown that now cavity 1 consistently has a much higher population than cavity 2.
		This imbalance is still present when sufficiently many long trajectories are considered (we average 1000 trajectories with duration $200/\gamma$).
		We see that the histogram has a large broad peak around $n_{\rm diff}\approx 5$ and is strongly asymmetric around $n_{\rm diff}=0$.
		The mean value  is $2.65$.

\end{document}